\documentstyle[snow209,epsfig,times]{article}
\def \lta {\mathrel{\vcenter
     {\hbox{$<$}\nointerlineskip\hbox{$\sim$}}}}
\def \gta {\mathrel{\vcenter
     {\hbox{$>$}\nointerlineskip\hbox{$\sim$}}}}
\begin{document}
%\begin{flushright}
%MADPH-96-971
%\end{flushright}
\title{Topcolor Models and Scalar Spectrum
\thanks{To appear in ``Proceedings of the 1996 DPF Summer Study
on New Directions for High Energy Physics''.}}
\author{Gustavo Burdman\\
\mbox{ }
{\it Department of Physics, University of Wisconsin, Madison, WI 53706}}

\maketitle

%% Get rid of page numbering
\thispagestyle{empty}\pagestyle{empty}

\begin{abstract} 
We review the motivation and main aspects of Topcolor models
with emphasis on the spectrum of relatively light scalars and 
pseudo-scalars.
\end{abstract}

\section{Dynamical Generation of \lowercase{$m_t$}.}
The generation of a large fermion mass like $m_t$ is 
an extremely difficult problem in  theories of dynamical 
Electroweak Symmetry Breaking (ESB).
For instance, in Technicolor theories an Extended Technicolor (ETC) 
interaction 
is required in order to obtain fermion masses. The interaction of the 
ETC gauge bosons with fermions and technifermions 
gives rise to fermion masses through terms  of the form
\begin{equation}
m_f=\frac{g_{ETC}^{2}}{M_{ETC}^{2}} \langle \bar{T}_L T_R\rangle
\label{etcmf}
\end{equation}  
where $M_{ETC}$ is the ETC gauge boson mass and 
$\langle\bar{T}_LT_R\rangle$ is the technifermion condensate. 
Thus, in order to generate the correct value of $m_t$, 
the ETC scale has to be
of ${\cal O}(1{\rm~TeV})$, which is uncomfortably low.
Several modifications of the dynamics within technicolor have 
been proposed in order to accommodate such a large fermion mass 
\cite{other_tc}. 
On the other hand, in top-condensation models, the large top-quark mass
is obtained from a $\langle\bar{t}_Lt_R\rangle$  arising  
as a consequence of a new gauge interaction, Topcolor, which couples 
strongly
to the top quark. Topcolor generates four-fermion interactions of the form
\begin{equation}
\frac{g^2}{\Lambda^2}\bar{\psi}_Lt_R \bar{t}_R\psi_L 
\label{ffermi}
\end{equation}
where $\psi$ is the $(t,b)$ $SU(2)$ doublet and $\Lambda$ is the typical
scale of the new interactions. 
If the coupling $g$ is strong enough the top condensation
occurs. The top chiral symmetry is 
spontaneously broken and a large $m_t$ is generated. 
This implies the presence of Goldstone bosons. Originally~\cite{bhl}, 
it was proposed that these were identified with the longitudinal components
of the electroweak gauge bosons so that Topcolor would also be fully
responsible for ESB. 
 With the ESB scale defined as $v\approx 246{\rm ~GeV}$, 
the decay constant of these top-pions is given by the
Pagels-Stokar formula
\begin{equation}
f_{\pi_t}^2\simeq\frac{N_c}{16\pi^2}m_t^2\ln{\frac{\Lambda^2}{m_t^2}}
\label{pagstok}
\end{equation}
with $N_c$ the number of colors. From (\ref{pagstok}) it can be seen that,
in 
order for $f_{\pi_t}=v/\sqrt{2}$ and $m_t$ to be 
close to the measured value, the Topcolor scale $\Lambda$ has to be 
extremely
large ($\simeq 10^{15}~\rm{GeV}$). This translates into an acute fine-tuning
of the coupling $g$ in (\ref{etcmf}), which has to be adjusted to  
the critical value with unnaturally high precision. 
One way to avoid this problem within Topcolor models is to give up 
the idea that ESB is fully driven by the $<\bar{t} t>$ condensate. 
For instance, 
a cutoff scale $\Lambda\simeq{\cal O}(1~\rm{TeV})$ gives a non fine-tuned
coupling (a few percent above critical), but a top-pion decay constant
of the order of $f_{\pi_t}\simeq (50-60)~\rm{GeV}$, which gives only 
small masses
to the $W$ and the $Z$. In this version of Topcolor~\cite{hill},
a separate mechanism must be invoked to generate most of the $W$
and $Z$ masses. 
This is the case
of Topcolor-assisted Technicolor. Most of $m_W$ and $m_Z$,
as well as {\em small} ($\simeq 1~\rm{GeV}$) quark masses come from 
the Technicolor sector. Thus a small portion of $m_t$ also comes from ETC
terms, but  most of the top-quark mass is dynamically
generated by
the Topcolor mechanism. The explicit ETC quark mass terms for top and bottom
turn the top-pions into massive pseudo-Goldstone  bosons. Their masses can 
be estimated in the fermion loop approximation to be ~\cite{hill}
\begin{equation}
m_{\pi_t}^2\simeq \frac{N_c}{8\pi^2}\frac{m_{\rm ETC} m_t}{f_{\pi_t}^2}
\Lambda^2 \label{tp_mass}
\end{equation}
where $m_{\rm ETC}$ is the effective value of the ETC  
quark masses. Although these are initially of the order of $1~\rm GeV$, 
the ETC top-quark mass receives large radiative enhancements from the 
Topcolor interactions \cite{hill}. Thus one can have top-pion masses in the 
range 
$m_{\pi_t}\simeq (100-300)~\rm{GeV}$. 
If $m_{\pi_t}<m_t$, then the top quark would primarily decay as
$t\rightarrow\pi_{t}^+ b$. The current CDF measurement of  
$Br(t\rightarrow W^+ b)$ implies $m_{\pi_t}>150\rm ~GeV$ at $68\%$
confidence level \cite{lane}. In what follows we will assume that
the Topcolor enhancement to the ETC top mass is enough to make
$m_{\pi_t}>m_t$. 

The existence of the top-pions ($\pi_t^\pm,\pi_t^0$) is an essential 
ingredient in the 
Topcolor scenario, regardless of the dynamics responsible for the 
ESB sector and the other quark masses.
The presence of other low-lying states depends on the details
of the model. A Topcolor model is greatly
specified by choosing a mechanism of isospin
breaking, which selects the top-quark direction for condensation, 
leaving the
bottom quark unaffected.
The complete anomaly-free fermion content is necessary in order to
know the scalar spectrum of the model. 
These low lying states have, in  most cases, masses 
well below the cutoff scale $\Lambda$, which points at them as 
possibly the first signal for Topcolor. 

\section{Topcolor Models and Scalar Spectrum}
In all models the Topcolor group contains an $SU(3)_1\times SU(3)_2$ which 
at an energy scale $\Lambda$ breaks down to ordinary $SU(3)_c$. 
The $SU(3)_1$ is assumed to interact strongly with the third generation 
quarks.
After Topcolor breaking there is, in addition to the massless gluons, 
an octet
of massive colored vector particles: the top-gluons.
At this stage, if the Topcolor coupling is above critical we would have 
both $t$ and $b$ condensation. To avoid the latter, the effective 
Topcolor interaction must be isospin breaking. We describe two 
typical scenarios to implement this aspect of the theory.

\subsection{Models with an additional $U(1)$}
 An effectively isospin breaking
interaction is obtained by
embedding two new $U(1)$ interactions in the 
weak hypercharge group such that
$U(1)_1\times U(1)_2\longrightarrow U(1)_Y$,
with the $U(1)_1$ strongly coupled to the third generation. This leaves an 
additional color-singlet massive vector boson, a $Z'$. Both the top-gluon
and the $Z'$ have masses of order $\Lambda$.  
After integrating out these heavy particles, the interesting effective 
four-fermion interactions that are induced have the form \cite{bbhk}
\begin{eqnarray}
{\cal L}=&&\frac{4\pi}{M_B^2}\left\{\left(\kappa+\frac{2\kappa_1}{9N_c}
\right)
\bar{\psi}_Lt_R\bar{t}_R\psi_L \right.\nonumber\\
&&+ \left.\left(\kappa-\frac{\kappa_1}{9N_c}\right)
\bar{\psi}_Lb_R\bar{b}_R\psi_L \right\}
\label{l_ffermi}
\end{eqnarray}
Here, $\kappa=(g_3^2/4\pi)\cot^2\theta$ and  
$\kappa_1=(g_1^2/4\pi)\cot^2\theta'$, with $g_3$ and $g_1$ the $QCD$ and 
$U(1)_Y$ couplings respectively. The angles $\theta$ and $\theta'$
characterize the embedding of the Topcolor and the $U(1)_1\times U(1)_2$
groups in $SU(3)_c$ and $U(1)_Y$. The requirement that $SU(3)_1$ and
$U(1)_1$ couple strongly to the third generation translates into
the conditions $\cot^2\theta\gg 1$, $\cot^2\theta'\gg 1$. 
The criticality condition
\begin{equation}
\kappa-\frac{\kappa_1}{9N_c} < \kappa_{\rm critical} < 
\kappa+\frac{2\kappa_1}{9N_c} \label{crit_con}
\end{equation}
must be satisfied in order to obtain $\langle\bar{t} t \rangle\neq 0$ and
$\langle\bar{b} b\rangle =0$. 
Constraints on the top-gluon sector come from $t$ production, as well as 
$t$ and $b$ dijet mass distributions at the Tevatron~\cite{harris}.
This specific model is also constrained by the effects of the $Z'$
on low energy data, both at the $Z$ pole \cite{chi_ter}
and at low energies through
FCNC \cite{gb,bbhk}. However, it is possible to accommodate all these 
constraints with 
a Topcolor scale $\Lambda\gta 1 {\rm ~TeV}$ and still not have a fine
tuning problem. For instance, in the most general case, a $2~{\rm TeV}$
top-gluon gives a Topcolor coupling $4\%$ above its critical value. 
Therefore, one can imagine a scenario where the Topcolor gauge bosons 
are at the few-$\rm TeV$ scale, making their direct detection difficult.
 In this scenario it is possible that
the effects of Topcolor dynamics will appear at lower
energy scales due to the presence of a relatively light
scalar spectrum.

The effective interactions of (\ref{l_ffermi}) can be written
in terms of two auxiliary scalar doublets $\phi_1$ and $\phi_2$. 
Their couplings to quarks are given by \cite{dk}
\begin{equation}
{\cal L}_{\rm\em eff.}=\lambda_1\bar{\psi}_L \phi_1 t_R +
\lambda_2\bar{\psi}_L \phi_2 b_R +{\rm h.c.}
\label{phi_qks}
\end{equation}
where $\lambda_1^2\equiv4\pi(\kappa+2\kappa_1/9N_c)$ and 
$\lambda_2^2\equiv 4\pi(\kappa-\kappa_1/9N_c)$.
At energies below $\Lambda$ the auxiliary fields acquire kinetic
terms, becoming physical degrees of freedom. 
With the properly renormalized fields $\phi_i^r=z^{1/2}_i\phi_i$
the criticality conditions (\ref{crit_con}) are equivalent to
\begin{equation}
\langle \phi_1^r \rangle = f_{\pi_t} \quad\quad \langle \phi_2^r \rangle= 0
\label{vevs}
\end{equation}
The $\phi_1$ doublet acquires a Vacuum Expectation Value (VEV)
giving mass to the top quark through
the coupling in (\ref{phi_qks}). It is of the form 
\begin{equation}
\phi_1^r=\left(\begin{array}{c}
f_{\pi_t} + \frac{1}{\sqrt{2}}\left(h_t+i\pi_t^0\right) \\
\pi_t^- 
\end{array}\right)
\label{phi_1} 
\end{equation}
As mentioned earlier, the set of three top-pions acquires 
a mass  from explicit small quark mass terms (the ETC
masses in Topcolor-assisted Technicolor). 
There is also a scalar, the $h_t$ or top-Higgs, which mass is estimated 
in the Nambu--Jona-Lasinio (NJL) approximation to be 
\begin{equation}
m_h\simeq 2m_t\label{mthiggs}
\end{equation}
The second doublet is present as long as 
the Topcolor interaction couples to $b_R$, as is the case in 
(\ref{l_ffermi}).
It is given by 
\begin{equation}
\phi_2^r=\left(\begin{array}{c}
H^+ \\
\frac{1}{\sqrt{2}}\left(H^0+iA^0\right) 
\end{array}\right)
\label{phi_2}
\end{equation}
These states are deeply bound by the Topcolor interactions and therefore 
can be light.  Their masses can be estimated once again within the NJL
approximation. For instance for $\kappa_1\simeq 1$ and 
$\Lambda=(2-3)~{\rm TeV}$ one has, for the neutral states \cite{dk}, 
\begin{equation}
m_{H,A}\simeq (150-330)~{\rm GeV}, \label{bpi_0} 
\end{equation}
whereas the mass of the charged states is determined by the 
relation 
\begin{equation}
m_{H^\pm}^2=m_{H,A}^2 + 2m_t^2 . \label{bpi_pm}
\end{equation}
The couplings to quarks can be read off equation 
(\ref{phi_qks}). Calculating 
the field renormalization constants $z_i$ to one loop and 
using (\ref{pagstok}), the couplings are simply $m_t/f_{\pi_t}$
where $m_t$ is the dynamically generated top quark mass. This is
a typical Goldberger-Treiman factor. Recalling that in Topcolor models
we expect $f_{\pi_t}/v \simeq 3$, we see that the coupling 
of the Topcolor ``Higgs'' sector to the top quark is considerably 
larger than that of the SM Higgs boson. 
Moreover, in models where the second doublet is present
this couples to $b$ quarks with the same strength as $(h_t,\pi_t^0)$
couple to top. This implies that $H^0$ and $A^0$ decays are dominated
by the $\bar{b}b$ final state.
The existence of these relatively light scalar states strongly coupled to 
third generation quarks implies a very rich phenomenology. 
In what follows we analyze the implications of the scalar spectrum
in the model described above. We discuss other alternatives in model
building in the next section, with emphasis on the differences in the 
scalar spectrum and phenomenology. 

{\em\bf Top-pions:} As discussed earlier, these are the pseudo-Goldstone
bosons of the breaking of the top chiral symmetry. They couple to 
the third generation quarks as
\begin{equation}
\frac{m_t}{\sqrt{2}f_{\pi_t}}\left( i\bar{t}\gamma_5 t \pi^0 +
i\bar{t}_R b_L \pi^+ + i\bar{b}_L t_R \pi^- \right) \label{top_pi}
\end{equation}
Although their masses are lifted by the ETC interactions, they can still
play an important role in low energy observables such as
as rare $B$ decay branching fractions and angular distributions
~\cite{gb,bbhk}, 
as well as in electroweak precision observables at the $Z$ pole
like $R_b$ \cite{bk}. 
Top-pions do not have two-gauge-boson couplings, and thus single 
$\pi_{t}$ production must involve a triangle diagram. At hadron colliders
they are mostly produced through the 
gluon-gluon-$\pi^0_t$ effective coupling induced by the top loop. 
For instance, the gluon-gluon fusion has a cross
section larger than that of the s-channel production of the 
SM Higgs boson by a factor of \cite{gbmp} 
\begin{equation}
r^2\equiv\left(\frac{v}{\sqrt{2}f_{\pi_t}}\right)^2.
\label{def_r}
\end{equation}
This enhancement is also present in the $\pi_t$ production in association
with top quarks, or in any production mechanism involving the 
$\bar{t}t\pi_t^0$ coupling. Production at $e^+e^-$ colliders
is discussed in \cite{gbmp}. 
 
If $m_{\pi_t}<2m_t$ the top-pion would be  narrow 
($\Gamma_{\pi_t}\lta 1{\rm~GeV}$). 
In models as the one presented above, where $b_R$ couples to 
to the strong $SU(3)_1$ interaction, instanton effects
induce a coupling of $\pi_t$ to $b_R$, which is  not 
present in (\ref{top_pi}). This implies that, as long as the $\bar t t$
channel is not open, the dominant decay mode of the $\pi_t^0$ is to 
$\bar b b$ \cite{gbmp}. Also, and independently of the presence of 
instanton effects, there is a coupling of $\bar c c$ to $\pi^0$ 
given by (\ref{top_pi}) times two powers of the $t\rightarrow c$ 
mixing factor arising from the rotation of weak to mass quark eigenstates. 
However, in the present model the $\bar b b$ mode is expected to dominate. 

{\em\bf Top-Higgs:} It corresponds to a loosely bound, CP even,  
$\bar t t $ state, coupling to the top quark with strength 
$m_t/(\sqrt{2}f_\pi)$. 
In the NJL approximation its mass is given by (\ref{mthiggs}),  
and therefore details of the non-perturbative 
dynamics are crucial to understand 
the decay modes and width of the $h_t$.   
The top-Higgs production is analogous to the $\pi_t$ case. 
The main difference between these two states is that $h_t$
couples to gauge boson pairs. These couplings are {\em suppressed}
with respect to the case of a SM Higgs by a factor of $1/r$.
However, if the $\bar{t} t$ channel is not open this would be 
the dominant decay mode. If this is the case, $\Gamma_{h_t}$ will
be considerably smaller than the width of a SM Higgs, whereas its
production cross section will still be $r^2$ times larger. 
On the other hand, if $m_{h_t}>2m_t$ the cross section is still 
the same but $\Gamma_{h_t}$ is $r^2$ {\em larger} than the width
of the SM Higgs of the same mass, so that $h_t$ could only  be 
detected as an excess in the $\bar t t $ in a given channel.  

{\em\bf The ``b-pions'' $H_b^0,A_b^0,H_b^\pm$:} If these states are
present (i.e. if Topcolor couples to $b_R$) they give  potentially dangerous
contributions to various low energy observables.
Their couplings to 
quarks are analogous to those in (\ref{top_pi}). The strongest constraint
on a model containing these bound-states comes from $B^0-\bar{B}^0$
mixing \cite{dk} and implies the existence of large suppression factors
 in the
quark mixing matrices~\cite{bbhk}. A constraint independent of
these details  is the contribution to $\Delta\rho_*=\alpha T$, 
the parameter measuring deviations from the SM $\rho$ parameter, 
due to the splitting between the neutral and charged states. 
%This takes the form
%\begin{equation}
%\Delta\rho_*=\frac{G_F}{8\sqrt{2}\pi^2} \left\{
%m_{H_b^\pm}^2+m_A^2 -2\frac{m_{H_b^\pm}^2 m_A^2}{m_{H_b^\pm}^2-m_A^2}
%\ln\frac{m_{H_b^\pm}^2}{m_A^2}\right\}
%\label{del_rho}
%\end{equation}
As an illustration of the size of the effect, we plot $\Delta\rho_*$
in Fig.~1 as a function of the mass of the neutral scalars
and making use of the NJL result (\ref{bpi_pm}) for the mass
splitting. The horizontal lines represent the $95\%$ c.l. interval
obtained using $\alpha_s(M_Z)=0.115$ \cite{cdt}. Although 
the bound is tighter as $\alpha_s(M_Z)$ increases, the b-pion splitting, 
always present in models with the additional $U(1)$'s, is not 
in contradiction
with electroweak precision measurements. 
\begin{figure}[b]
\leavevmode
\centering
%\epsfysize=5cm
%\epsfbox{prueba.ps,angle=180}
\epsfig{file=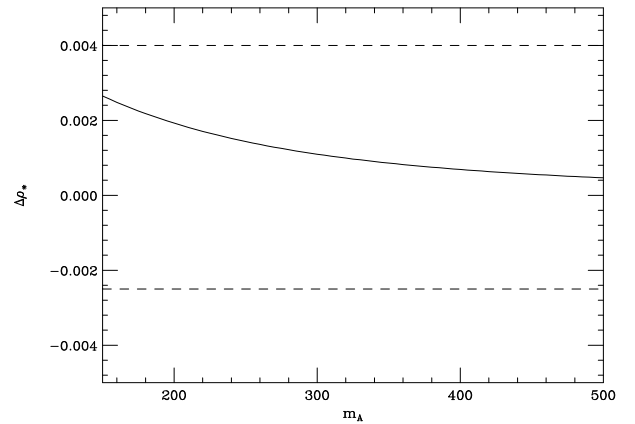,width=8cm,angle=90}
\caption{The contribution to $\Delta\rho_*$ due to the 
mass splitting among b-pions.}
%
%\label{fig:layout}
%
\end{figure}

These states decay almost exclusively to $b$ pairs and they are extremely 
broad. Their production proceeds in similar ways to that of $h_t$ 
and $\pi_t$,
with the important difference that the quark inside the triangle loop, 
the $b$
quark, is much lighter than $\sqrt{s}$ which translates into an effective
suppression of the amplitude from its ``hard'' value of $m_t/f_{\pi_t}$.

In addition to these low lying scalar states, the Topcolor interactions
in principle lead to the formation of heavier bound states.
For instance, 
there will be a color singlet vector meson, the top-rho $\rho_t$. 
However, $\rho_t$ not only couples to top-pions but also couples
directly to third generation quarks and with strength proportional
to $m_t/f_{\pi_t}$. Moreover, their mass can be estimated in the NJL
approximation to be of the order of the cutoff $\Lambda$ \cite{bk}. 
This suggests that the influence  of $\rho_t$ in low energy
observables as well as in production of Topcolor bound-states 
(e.g. top-pions through vector meson dominance) is largely
suppressed.

\subsection{``Axial'' Topcolor}
In this type of Topcolor models, the strong $SU(3)_1$ group does not couple
to $b_R$, barring the possibility of a $\langle\bar{b}_L b_R\rangle$
condensate. An example was presented in \cite{bbhk}. The cancelation of
 anomalies
requires the introduction of a new set of fermion fields, $Q^{a}_{L,R}$,
 with $a=1,..., N_Q$. There is a new interaction, $SU(N_Q)$. 
The novelty is that, depending on the choice of $N_Q$, light
 quarks might have
to ``feel'' the strong Topcolor interaction. For instance for  $N_Q=3$
the fermions must transform under $SU(3)_Q\times~SU(3)_1\times~SU(3)_2$ as
\begin{eqnarray}
(t,b)_L~ (c,s)_L & \simeq& (1,3,1) \nonumber\\ 
t_R &\simeq & (1,3,1) \nonumber\\
Q_R &\simeq & (3,3,1) \nonumber\\
(u,d)_L &\simeq & (1,1,3) \nonumber\\
(u,d)_R ~(c,s)_R &\simeq & (1,1,3) \nonumber\\
b_R &\simeq & (1,1,3)\nonumber\\
Q_L &\simeq & (3,1,3). \nonumber
\end{eqnarray}
As advertised above, $b_R$ is not coupled to the strong Topcolor 
interaction, 
whereas the cancelation of anomalies now requires that $(c,s)_L$ transforms
as a triplet under $SU(3)_1$. 
The quarks have standard $U(1)_Y$ assignments, whereas the $Q_{L,R}$ have 
$Y=0$. Furthermore, they are $SU(2)_L$ singlets and therefore 
electrically neutral. If leptons are incorporated with their standard
$SU(2)_\times U(1)_Y$ quantum numbers and as singlets under 
$SU(3)_Q\times SU(3)_1\times SU(3)_2$, all anomalies cancel.  
The $SU(3)_Q$ forms a $\langle\bar{Q}Q\rangle$ condensate which, in turn, 
breaks Topcolor down to $SU(3)_c$ dynamically. 

The $SU(3)_1$ is chiral-critical and leads to the formation of a 
$\bar{t} t$ condensate and a dynamical $m_t$. 
This breaks an $SU(4)_L\times U(1)_L\times U(1)_R$ global chiral 
symmetry and leads to the existence of a composite
scalar field $F$, quadruplet under $SU(4)_L$. This can be decomposed 
into two doublets, one of which is just $\phi_1^r$ of Section II.A, 
containing
$h_t$ and the top-pions. The additional doublet, $\cal C$, contains a 
set of three pseudo-scalars, the ``charm-top-pions'' or 
$\vec\pi_c$, as well as a ``charm-top-Higgs'' $h_c$:
\begin{equation}
\cal C = \left(\begin{array}{c}
\frac{1}{\sqrt{2}}(h_c+i\pi_c) \\
\pi_c^- 
\end{array}
\right) 
\label{c_pions}
\end{equation}
The masses of the charm-top-pions are generated by the same 
terms that break chiral symmetry explicitly and  induce the top-pion
masses, so they are expected to be of the same order as $m_{\pi_t}$. 
The mass of the $h_c$ in the NJL approximation is $\approx m_t$.
The couplings to quarks are analogous to those of the top-pions 
and have the
form \cite{bbhk}
\begin{equation}
\frac{m_t}{f_{\pi_t}}~(\bar{c}~\bar{s})_L~{\cal{C}}~t_R  + {\rm h.c.}
\label{ctp_toq}
\end{equation}
Important constraints on this scenario come from low energy observables. 
For instance, $D^0-\bar{D}^0$ mixing is mediated by three-level s-channel
exchange of $h_c$ and $\pi_c^0$. However, the mixing amplitude is
proportional to the product of the various unknown quark rotation
factors entering the transformation from the weak to the mass eigen-basis
for quarks \cite{bbhk}. These factors depend on the sector of the theory
responsible for light quark masses, e.g. ETC. Thus, charm mixing is a direct
constraint on the light-quark mass matrices generated by this sector. 
On the other hand, charm-top-pion contributions to $R_c$ and $R_s$ 
are potentially large \cite{bk} and independent of these details. 

The production of top-pions in this model is completely analogous to the
model in the prevoius section. However, here the top-gluons do not couple 
to $b_R$, so there will be no 
instanton induced $b$ quark mass term and therefore no 
$\pi_t^0\rightarrow \bar b b$ decay mode. 
Thus, if top-pions have a mass below the $\bar t t$ threshold, 
the neutral states will primarily decay to gluons. 
Also and as we mentioned in the previous model, 
the $\bar c c$  decay mode 
is suppressed by two factors coming from the $t\rightarrow c$
rotations. Although these factors are not very constrained,  
naive estimates lead to values of 
$\Gamma(\pi^0_c\rightarrow\bar{c} c)$ that 
indicate that this mode 
is competitive with the gluon-gluon channel \cite{gbmp}.  

The production of charm-top-pions does not proceed via gluon-gluon
fusion. Production through the anomaly is no only suppressed by
loop factors and couplings but also by CKM factors. The most efficient
mechanism for single production of $h_c$ or $\pi_c^{0,\pm}$ 
is by  emission off a top quark line, with this turning into a 
charm or strange quark for neutral and charged emission respectively.
The multijet final state, e.g. $t\bar t c\bar c$ for the neutral case, 
may be difficult to  separate from the QCD background, especially given that
these scalars are expected to be very broad. More detailed studies 
are necessary.  
Charm-top-pions can also be pair produced, just like 
the  top-pions and the top-Higgs,  via their 
model independent couplings to gauge bosons. However, as in the case
of the previous section, these couplings are also suppressed 
by the ratio $f_{\pi_t}/v$. 

It is possible to extend this model to the first family by just including
the $(u, d)_L$  among the quarks strongly coupled to the Topcolor
interactions, in addition to $(c, s)_L$. 
Such  Topcolor models are partly motivated by the possibility
of the existence of an excess of events at high energies in the 
CDF data for the inclusive jet cross section \cite{cdf_incl}.
In Ref.~\cite{sc_es} the consequences of 
a universally coupled top-gluon on the inclusive jet production
were studied. 
It is possible to write down an anomaly free realization
of this proposal.  
The only other modification
needed to insure the cancelation of all anomalies is that now we 
need $N_Q=5$. 
Thus, $(u,d)_L$ transforms as $(1,3,1)$ under 
$SU(5)_Q\times SU(3)_1\times SU(3)_2$, and the other fermions 
transform as before, with the exception of the $Q_{L,R}$ that 
transform as quintuplets under $SU(5)_Q$. 
The global chiral symmetry broken by the dynamical top quark mass
is now $SU(6)_L\times U(1)_R$. 
Therefore, besides the scalar content of the two previous models,  
there is a new scalar doublet, $\cal U$, the ``up-top-pions''. 
Their masses are very similar to those of $h_t$ and $m_{\pi_c}$.
Their couplings to quarks can be read off (\ref{ctp_toq}), with the
replacements $\cal C \rightarrow\cal U$ and 
$(\bar{c}~\bar{s})_L\rightarrow(\bar{u}~\bar{d})_L$. 
Although these couplings imply an additional contribution
to $D^0-\bar{D}^0$ mixing, it is governed by the same product
of up-quark rotation matrix elements, and therefore can be 
avoided by the same choice of mass matrices that suppressed
the charm-top-pion contributions. 
There will also be new contributions to the hadronic $Z$ width \cite{bk}.
The production of $h_u$ and $\pi_u$ has the same features as that of the
charm-top-pions. In both cases, one expects these resonances to be very 
broad and a careful study is needed in order to establish their 
detectability in the various channels and in different environments.

\section{Conclusions}
We have reviewed the main aspects of Topcolor models and the constraints 
derived
from the spectrum of low laying scalars and pseudo-scalars. 
A basic feature of all Topcolor models is the existence of the loosely bound 
state
$h_t$ and a triplet of top-pions $\vec{\pi_t}$. The latter is present 
in the physical 
spectrum in models where Topcolor is not solely responsible for ESB 
(e.g. Topcolor-assisted
Technicolor). 

Although the production of $h_t$ shares several features with that of the
SM Higgs, it has a larger cross section and a very different width
(larger or smaller by $r^2$ depending on $m_{h_t}$).
On the other hand, the observation of  $\pi_t^0$  at hadron 
colliders is problematic given that its decay modes, 
$\bar b b$, $\bar c c$ and gluon-gluon,  
are very hard to 
extract from the background. 
Its production at lepton colliders is discussed in \cite{gbmp}. 

Specific realizations of Topcolor tend to have additional, relatively light, 
scalars. 
Given that these tend to be very broad objects, their detectability 
at various
facilities requires a careful study, particularly at hadron colliders. 
Considering that Topcolor theories are still at relatively early
model building stages, 
the study of low energy signals of and constraints on the scalar spectrum 
in the various models will play a central role in determining 
what the dominant 
Topcolor phenomenology will be in  future experiments.

%Figures and equations are numbered with arabic 
%numerals, while tables are numbered with uppercase roman 
%numerals, as in Table~\ref{tab:}
%%%%% A Table
%%
%\begin{table}[h]
%\begin{center}
%\caption{Sample table in correct format.}
%\label{tab:sample}
%\begin{tabular}{lc}
%
%\hline
%\hline
%Element Type  & Numeral Type \\
%\hline
%Figure  &   1, 2, 3, etc.  \\
%Equation &  1, 2, 3, etc.  \\
%Table   &   I, II, III, etc.  \\
%\hline
%\hline
%
%\end{tabular}
%\end{center}
%\end{table}
%%
%%%%%

%Figure~\ref{fig:layout} shows the correct layout of your document, 
%and is an example of figure formatting style.

%%%%% A Figure
%%
%
%\begin{figure}[b]
%
%\leavevmode
%
%\centering
%\epsfysize=5cm
%\epsfbox{prueba.ps,angle=180}
%\epsfig{file=prueba.ps,width=5cm,angle=90}
%
%\caption{The contribution to $\Delta\rho_*$ due to the 
%splitting among the b-pions.}
%
%\label{fig:layout}
%
%\end{figure}
%%
%%%%%

 {\bf ACKNOWLEDGMENTS:} Much of this work is based on 
collaborations
and/or conversations with Sekhar Chivukula, Chris Hill, Dimitris Kominis
and Elizabeth Simmons. The author thanks Michael Peskin for pointing
out the possible effects of b-pions in $\Delta\rho_*$. The first part
of this work was done while the author was at the Fermilab Theory
Group. 
This research was supported in part by the U.S.~Department of Energy under  
Grant No.~DE-FG02-95ER40896 and in part by the University of 
Wisconsin Research  
Committee with funds granted by the Wisconsin Alumni Research Foundation.

%%%%% References
%

\end{document}